\documentclass[a4paper,11pt]{article}
\usepackage{pos}

\title{Testing of KNO-scaling of charged hadron multiplicities within a Machine Learning based approach}
\ShortTitle{Testing of KNO-scaling of charged hadron multiplicities within a ML based approach}

\author*[a]{Gábor Bíró}
\author[a,b]{Bence Tankó-Bartalis}
\author[a]{Gergely Gábor Barnaföldi}

\affiliation[a]{Wigner Research Center for Physics,\\
  29--33 Konkoly--Thege Mikl\'os Str., H-1121 Budapest, Hungary.}
\affiliation[b]{University of Oxford, University Offices, Wellington Square, Oxford, OX1 2JD, United Kingdom.}

\emailAdd{biro.gabor@wigner.hu}
\emailAdd{tanko.bartalis.bence@wigner.hu}
\emailAdd{barnafoldi.gergely@wigner.hu}

\abstract{
The results of a Machine Learning-based method is presented here to investigate the scaling properties of the final state charged hadron and mean jet multiplicity distributions. Deep residual neural network architectures with different complexities are utilized to predict the final state multiplicity distribution from the parton-level final state, generated by the \textsc{Pythia} Monte Carlo event generator. Hadronization networks were trained by $\sqrt{s}=7$~TeV events, while predictions have been made for various LHC energies from $\sqrt{s}=0.9$~TeV to 13~TeV. Scaling properties were adopted by the networks at hadronic level, indeed KNO-scaling is preserved---although, the scaling of the mean jet multiplicity distributions varies for the applied models.
}

\FullConference{%
  41th International Conference on High Energy Physics, \\
  6 - 13 July, 2022\\
  Bologna, Italy
}

\renewcommand{\logo}{\relax}

\begin{document}
\maketitle

\section{Introduction}






The utilization of Machine Learning techniques in high-energy physics is getting more-and-more popular, with increasing number of successful applications~\cite{Feickert:2021ajf, chollet2015keras, abadi2016tensorflow}. Neural networks (NN) perform well, representing complex phenomenological models of the high-energy collisions---see \cite{Biro:2021zgm, Mallick:2022alr, Monk:2018zsb} and the further references therein. Modelling the jet formation and/or hadronization is especially a good testbed by applying deep residual networks. Therefore, numerous studies investigate and calculate the most fundamental experimental observables in proton-proton collisions~\cite{Biro:2021zgm, Mallick:2022alr, Monk:2018zsb, he2015deep}. A primary result is the generalization capability of the NN models: although they were trained at a single center-of-mass energy of $\sqrt{s}=7$ TeV, trained networks carry information on a wide kinematical ranges and multiplicities at other energies as well. Therefore, complex architectures are able to predict qualitatively good observational values at any LHC energies.

In high-energy proton-proton collisions, the mean (charged) hadron multiplicity, $P_n$ depends on the center-of-mass energy. Moreover, with appropriate reformulation, the multiplicity distributions of charged hadrons and jets collapse onto a universal scaling curve at all energies. This observation is known as the KNO-scaling~\cite{KOBA1972317, Hegyi:2000sp, Vertesi:2020utz}: $P_n=\frac{1}{\left<n\right>}\Psi\left(\frac{n}{\left<n\right>}\right)$, which is violated at higher energies~\cite{CMS:2010qvf}.

In the current study we investigate the KNO-scaling for multiplicity distributions of charged hadrons and jets at proton-proton collisions. Results were performed by neural networks at various LHC energies from $\sqrt{s}=0.9$ to $13$~TeV.

\section{Results from the Model}

\begin{figure}[htb]
\centering
\includegraphics[width=0.49\linewidth]{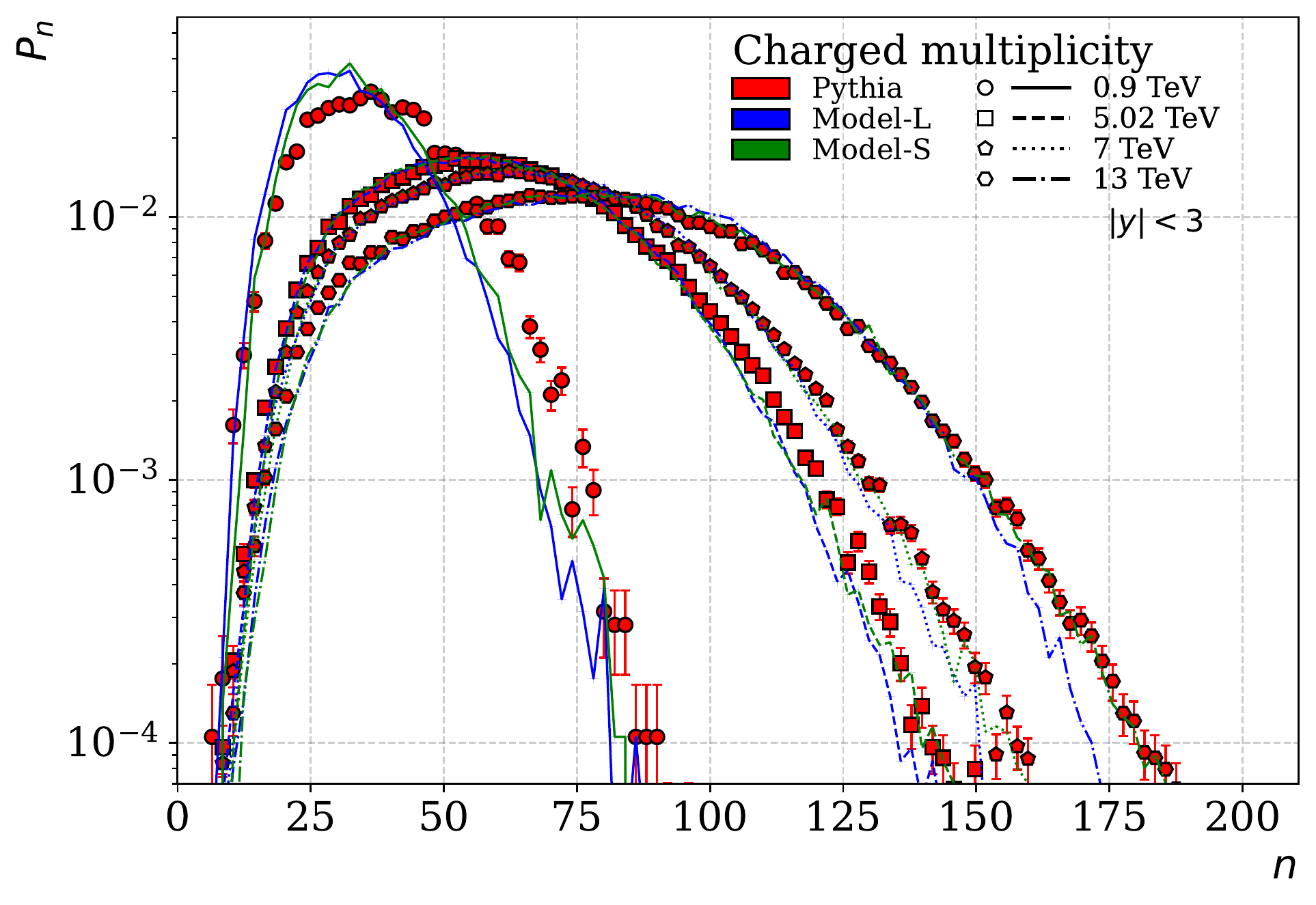}
\includegraphics[width=0.49\linewidth]{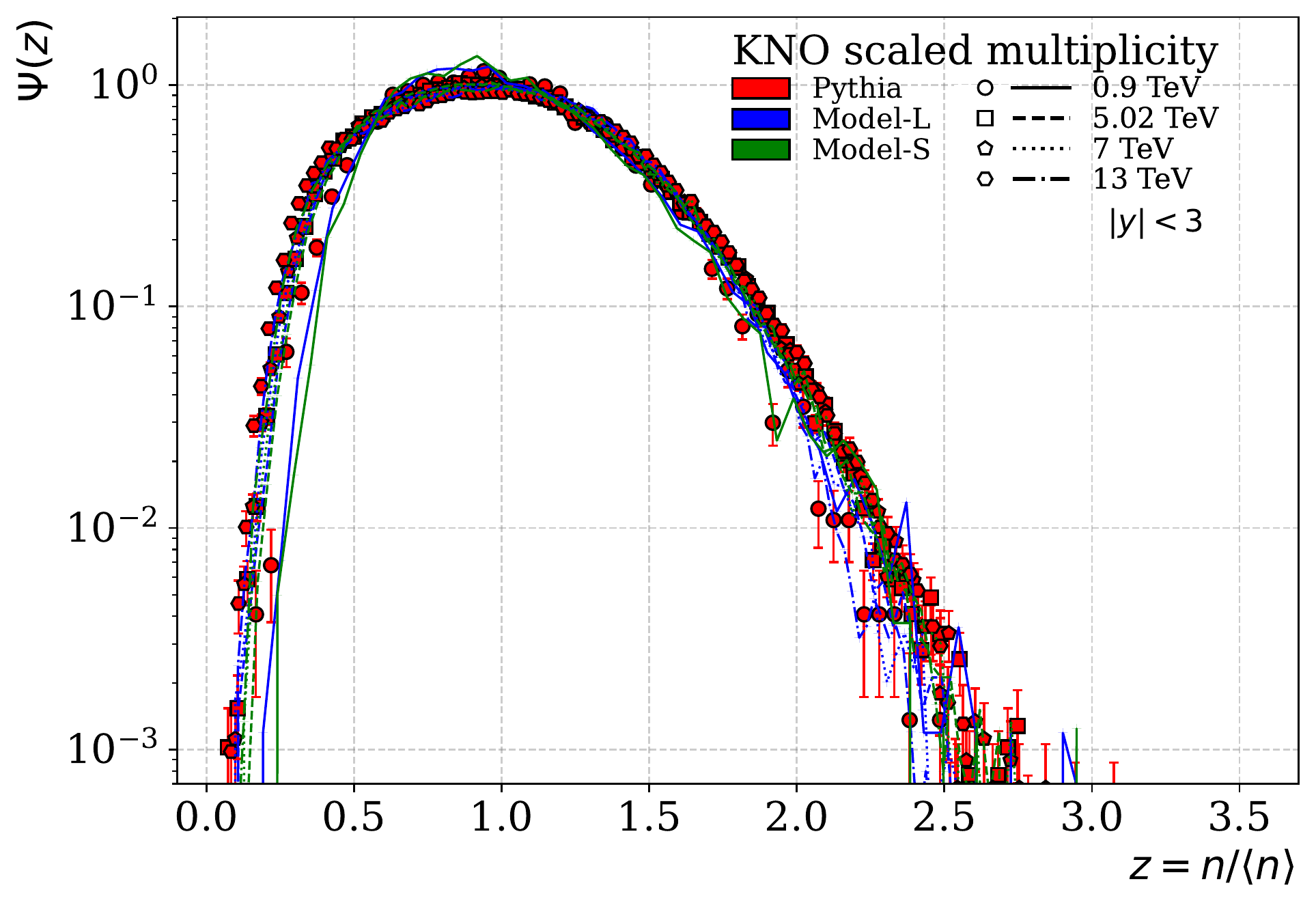}
\caption{Charged hadron multiplicity and the KNO-scaled multiplicity in proton-proton collisions, $|y|<3$.}
\label{fig:mult1}
\end{figure}

The multiplicity distributions of charged hadrons are presented here. We calculated partonic final states by the \textsc{Pythia} Monte Carlo generator  and applied a neural network based 'hadronization' algorithm, instead of the Lund hadronization model in \textsc{Pythia}~\cite{Sjostrand:1982fn, Sjostrand:2014zea, Cacciari:2011ma}. Results are shown in the left panel of Figure~\ref{fig:mult1}, where the \textsc{Pythia}-calculated multiplicity distributions are compared with the predicted results for various c.m. energies in a wide rapidity region of $|y|<3.0$. The right-hand side plot shows the corresponding $\Psi(z)$ scaling functions, with the joint curves as the effect of KNO-scaling. Note that, mostly, jetty events  are considered (i.e. events where at least 2 jets with $p_{T_J}\geq40$ GeV and $R=0.4$ are present). See Ref.~\cite{Biro:2021zgm} for the event selection criteria and for the specific model details.



We found, the Machine Learning approach results in a good agreement with the reference Monte Carlo model at all LHC energies, indicating that the neural networks were able to adopt the multiplicity scaling. Moreover, the KNO-scaling seems to be fulfilled for the models, presenting only slight deviations due to statistical reasons and especially at the lowest energy and at the highest multiplicities. According to the scaling curves of Figure~\ref{fig:mult1}, the higher event multiplicity does not eventuate in a significant scaling violation. This observation is opposite to the measurement by the CMS Collaboration, however, in a less biased data sample~\cite{CMS:2010qvf}.

\begin{figure}[htb]
\centering
\includegraphics[width=0.49\linewidth]{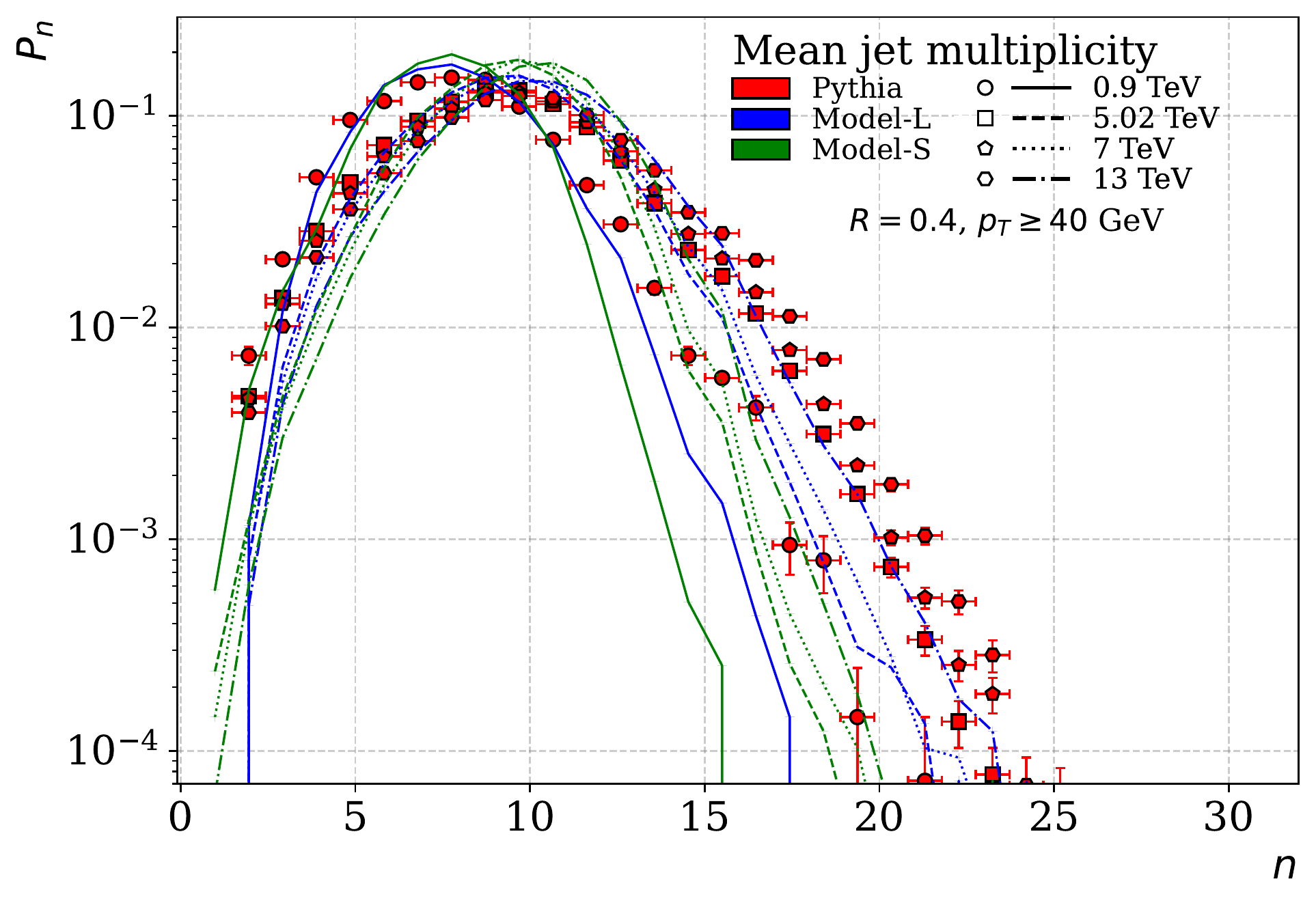}
\includegraphics[width=0.49\linewidth]{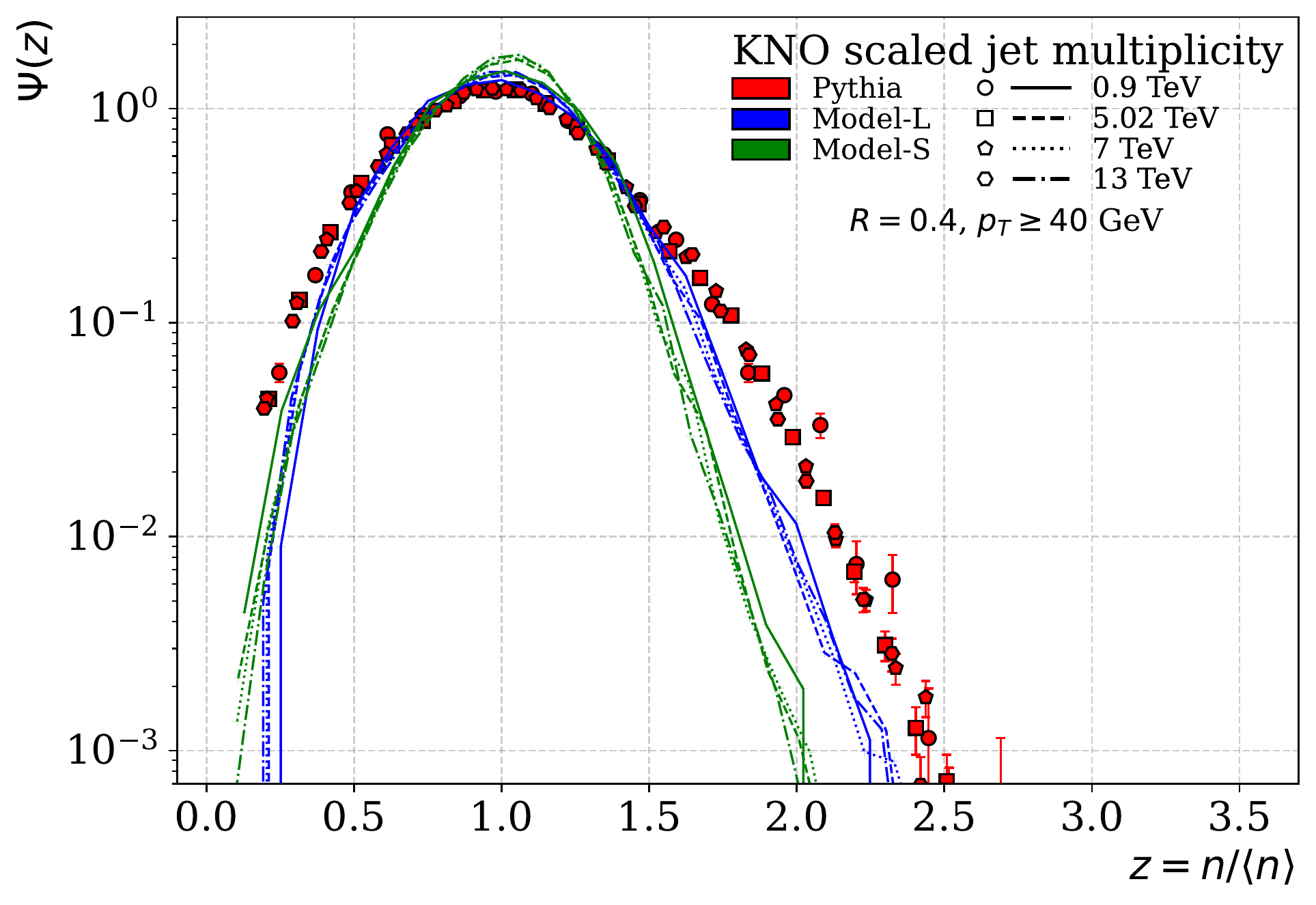}
\caption{Jet mean multiplicity and the KNO-scaled multiplicity in proton-proton collisions, for jets with $p_{T_J}\geq40$ GeV and $R=0.4$.}
\label{fig:mult3}
\end{figure}

Figure~\ref{fig:mult3}. shows the mean jet multiplicity distributions (left panel) and the their KNO-scaled curves (right panel). Although, due to statistical reasons the neural network based predictions present a larger deviation with respect to the reference \textsc{Pythia} results, the scaling manifests in this case, as well. The scaled curves collapse into a universal curve at all energies for each  investigated model, but the predicted scaling itself is differently grouped. This is an interesting outcome, which might be originated in different entropy production mechanisms of the models.

\section{Summary}

In this contribution the scaling properties of neural network based hadronization model calculations of charged hadron multiplicities and jets at LHC energies were presented. The neural networks were able to adopt the KNO-scaling in jetty events at $|y|<3$ rapidity. Although the training of the neural networks were done at $\sqrt{s}=7$ TeV energy in proton-proton collisions, predictions also for various other LHC-energies resulted in a good agreement with the reference \textsc{Pythia} calculations. On the other hand, the mean jet multiplicity distributions revealed diverse scaling behavior for the different models.

\section*{Acknowledgements}
The research was supported by the Hungarian National Research, Development and Innovation Office OTKA K135515, 2019-2.1.11-T\'ET-2019-00078, 2019-2.1.11-T\'ET-2019-00050, 2020-2.1.1-ED-2021-00179 
and 2022-1.2.3-GYAK-2022-00014 grants and by the Wigner Scientific Computing Laboratory. Author G.B. was supported by the EU project RRF-2.3.1-21-2022-00004 within the framework of the Artificial Intelligence National Laboratory.

\bibliographystyle{JHEP}
\bibliography{papers_all}

\end{document}